\documentclass[12pt]{article} 
\font\fff=eufm10 scaled \magstep1
\def\g{\hbox{\fff g}}
\begingroup 
\newtheorem{thm}{Theorem}[section]   

\newtheorem{proposition}[thm]{Proposition}
\newtheorem{theorem}[thm]{Theorem}

\newtheorem{definition}[thm]{Definition}
\newtheorem{notation}{Notation}

\newtheorem{remark}[thm]{Remark}            
\newtheorem{observation}[thm]{Observation}  
\endgroup
\def\state#1.{|#1\rangle}
\def\qed{\qquad\framebox[7pt]\medskip\noindent}

\def\half{{\textstyle\frac{1}{2}}}
\newenvironment{proof}{{\sl Proof:}\quad}{\hfill{\qed}\\ \noindent}
\renewenvironment{abstract}{\begin{quote}{\bf Abstract.\
}\small}{\end{quote}\bigskip}
\title{Canonical variables and analysis on so(n,2)}
\author{Philip Feinsilver${}^*$ and Jerzy Kocik\\ 
\emph{ Department of Mathematics}\\
\emph{ Southern Illinois University}\\
\emph{ Carbondale, IL. 62901, U.S.A.} \\
\strut\\
Michael Giering\footnote{A talk based on an earlier
  version of this work was presented (P.F.,M.G) in the special session on
  Algebraic Methods in Statistics organized by G. Letac at the 1997 AMS
  meeting in Montreal.}\\ 
\emph{CDS, Information Services International}\\
 \emph{100 International Drive}\\
\emph{ Mt. Olive, NJ 07828}
}
\date{}

\begin{document}

\maketitle

\begin{abstract}
The approach of Berezin to the quantization of so($n$,2) 
via generalized coherent states is considered
in detail. A family of $n$ commuting observables is found in which the
basis for an associated Fock-type representation space is expressed.
An interesting feature is that computations can be done by
explicit matrix calculations in a particular basis. The basic
technical tool is the Leibniz function, the inner product of coherent states.
\end{abstract}

\vfill
\pagebreak

\section{Introduction}
In the papers
\cite{BZ74, BZ75a, BZ75b}, Berezin presents an 
approach to quantization using
generalized coherent states, as explained by Perelomov \cite{PLV},
also see \cite{BaGi} and the survey \cite{KS}. 
\\

We first recall the Cartan decomposition and relate it
to Berezin's theory. In the section following, we give the matrix version
used for basic computations. The representation space is
constructed. Then the Leibniz function is computed. The observables,
the natural variables for analysis on so($n$,2), are found and their
joint spectral density is discussed. We conclude showing how the Lie
algebra is recovered from the Leibniz function.
\\

Work most closely related to this paper is that of Onofri \cite{Of}
and Berceanu \& Gheorghe \cite{BeGh}.  The coherent state methods 
given by Hecht \cite{He} are closely related to the present article as
well.
\\

In addition to the works cited above, in the mathematics literature
we have found the book by Hua \cite{Hua} and 
the paper of Wolf \cite{Wolf} very useful. An exposition of 
the present authors' theory with emphasis
on connections with probability is given in \cite{FSVol3}.
A major aspect of the mathematical point of view is the theory of
symmetric cones and Jordan algebras. See \cite{Faraut-Koranyi} for
analysis in that context.
\\

The significance of the pseudo-Euclidean group SO($n$,2) in physics 
is well-known.  
It serves as a ``linearization'' of the conformal group 
of Minkowski space ${\bf R}^{n-1,1}$ (see e.g., \cite{Fel}), 
the symmetry group of Maxwell's equations.  
Also, the group SO($n$,2) plays an important r\^ole in the $n$-dimensional 
Kepler problem, where the compactified phase space (the Moser phase space) 
coincides with a coadjoint orbit of the dynamical group SO($n$+1,2) 
\cite{Kep-3, Kep-2, Kep-1}. 
In another context, the group SO(4,2) serves as the spectrum-generating symmetry group of 
the hydrogen atom  \cite{BaKl, MaMa}. 

\begin{remark}\rm
This paper is based on the Ph. D. dissertation of
the middle author (M.G.) \cite{MG}.
\end{remark}

{\it Note: matrix computations have been done using Maple V.}
\section{Cartan decomposition and Berezin theory}

Consider a Lie algebra $\g$.  At the heart of our construction is 
the existence of two abelian subalgebras $\cal R$ and $\cal L$ of the 
same dimension $n$, such that the Lie algebra they generate is $\g$
itself:
$\g = {\rm gen}\,\{\mathcal L, \mathcal R \}$. 
\\ 

An important case of such a structure is the 
\emph{Cartan decomposition} for symmetric Lie algebras, where $\g$ 
has the form 

\begin{equation}\label{eq:CC}
       \g = \mathcal{L}\oplus\mathcal{K}\oplus\mathcal{R}
\end{equation}

with $\mathcal{L}$ and $\mathcal{R}$ satisfying 
$[\mathcal{L},\mathcal{R}]\subseteq\mathcal{K}$, 
$[\mathcal{K},\mathcal{R}]\subseteq\mathcal{R}$, and 
$[\mathcal{K},\mathcal{L}]\subseteq\mathcal{L}$. 
\\

Denote bases for $\mathcal{R}$, $\mathcal{L}$ and $\mathcal{K}$ 
by $\{R_j\}$, $\{L_j\}$, and $\{\rho_A\}_{1\le A\le m}$, respectively.
\\

\begin{remark}\rm
Later in the paper we will give a Cartan decomposition of so($n$,2).  
The $\rho$ elements will be taken as generators of rotations in 
the purely spatial or temporal sectors of ${\bf R}^{n,2}$, while the 
$\mathcal{L}$ and $\mathcal{R}$ elements will be certain combinations of
boosts.
\end{remark}

A typical element $X\in\g$ is of the form 

\begin{equation}
\label{eq:CC2}
      X= v'_jR_j  + u'_A\rho_{A}+ w'_jL_j
\end{equation}
for some $(2n+m)$-tuple $(v',u',w')$.
We can express exponentiation of $X$ to an element of the group either by
the standard 
exponential map, or via factorization into subgroups corresponding to 
the decomposition of the Lie algebra, thus

\begin{equation}
\label{eq:grpel}
   e^X=\exp( v_iR_i) 
       \left( \prod_{A}\exp(u_{\dot A}\rho_{\dot A})\right)\exp(w_jL_j)
\end{equation}

(We use the convention of summation over repeated indices, unless they
are dotted; there is no summation over $\dot A$ above).
Clearly, the coordinates $(v,u,w)$ versus $(v',u',w')$ are mutually
dependent as they represent in (\ref{eq:grpel}) the same group element.\\

Our general goal is to construct a representation space for the enveloping 
algebra of $\g$ and then find an abelian subalgebra of self-adjoint 
operators to take as our observables of interest. \\

First, let us construct a Hilbert space $\mathcal{H}$ spanned by a 
basis 

\begin{equation}
\label{eq:fockbasis}
        \state k_1,k_2,\ldots,k_n. = 
             R_1^{k_1}\cdots R_n^{k_n}\Omega
\end{equation}
 
where $\Omega$ is a vacuum state. Define the action of the algebra elements  
on the vacuum state thus \\

(i)   \qquad  $\hat R_j   \Omega = R_j \Omega $       \\
(ii)  \qquad  $\hat L_j   \Omega = 0           $      \\
(iii) \qquad  $\hat \rho_A\Omega =\tau_A\Omega  $     \\

where $\tau_A$ are constants.
Next, assume that $\mathcal{H}$ admits a symmetric scalar product
(not necessarily hermitian!) in some number field, such that
the ladder operators are mutually adjoint with respect to it:
$$
         \hat R_i^*= \hat L_i
$$
Thus, there is a 1-1 map $\mathcal R\leftrightarrow \mathcal L$ that admits 
such a pairing via adjoints.  Additionally, we shall always consider the vacuum
state normalized, $\langle \Omega,\Omega\rangle=1$. \\

For the purpose of this paper, we shall assume that only one element 
of $\mathcal K$, say $\rho_0$, acts on $\Omega$ as a nonzero constant $\tau$, 
so that the group element specified by equation (\ref{eq:grpel}) acts on
$\Omega$ 
as follows 

\begin{equation}
\label{eq:grpaction}
        e^X\Omega = e^{\tau \,u}\,\exp(v_jR_j)\Omega 
\end{equation}

The system possesses two types of lowering and raising operators.
The \emph{algebraic lowering} and \emph{raising} operators are
defined simply by concatenation within the enveloping algebra
(operator algebra generated by the representation) of $\g$
followed by acting on $\Omega$, that is

\begin{eqnarray*}
 {\hat R}_j \psi &=& R_j \psi  \\
 {\hat L}_j \psi &=& L_j \psi
\end{eqnarray*}
for any linear combination $\psi$ of basis elements (\ref{eq:fockbasis}).
The ``hat" can be thus omitted without causing confusion.
We shall also introduce \emph{combinatorial raising operators}, ${\cal R}_j$, 
and  \emph{combinatorial lowering operators}, ${\cal V}_j$, acting on the
basis 
as follows

\begin{eqnarray*}
       {\cal R}_j\;\state k_1,k_2,\ldots,k_n. 
              &=& \state k_1,k_2,\ldots,k_j+1,\ldots k_n.\\
       {\cal V}_j\;\state k_1,k_2,\ldots,k_n. 
              &=& k_j\,\state k_1,k_2,\ldots,k_j-1,\ldots k_n.
\end{eqnarray*}

Notice that the operator ${\cal V}_j$ acts formally as the operator of partial
differentiation with respect to the corresponding variable $R_j$.\\

The algebraic raising operators are represented directly by the $\cal R$'s,
namely $\hat R_j={\cal R}_j$.
But the combinatorial lowering operators do not 
necessarily correspond to elements of $\g$.
The idea will be to express  the \emph{algebraic lowering operators},
$\hat L_j$,
(and hence the basis for $\g$), also in terms of
the operators $\{{\cal R}_j,{\cal V}_j\}$. 
\\

Let us introduce the coherent states as the image of the subgroup 
generated by the (abelian) subalgebra $\mathcal R\subset \g$ 
in the Hilbert space $\mathcal H$ constructed above, namely
$$ 
     \psi_v=\exp({v_jR_j})\Omega
$$
Thus, the coherent states are parametrized by the elements $v_jR_j$
of $\mathcal R$, or, equivalently, by coordinates $v=(v_1,\ldots, v_n)$.  
We shall denote the manifold of coherent states as 
$\mathcal C$ with the parametrization $\cal R\to C$.

\begin{observation}\label{rem:hatrep}\rm
When restricted to coherent states, $\mathcal R_{j}$ acts as differentiation 
$\partial/\partial v_j$, while ${\cal V}_j$ acts as multiplication by $v_j$. 
Hence, we can determine the action of any operator defined as a (formal) 
operator function $f(\mathcal R, \mathcal V)$, with all $\mathcal V$'s to the
right of any ${\mathcal R}_j$, by 
(1) moving all ${\cal R}$'s to the right of all ${\cal V}$'s in the
formula $f$, yielding the operator  $\check f(\mathcal R, \mathcal V)$,
and then 
(2) replacing $\mathcal{V}_j\to v_j$ and $\mathcal{R}_j\to
\partial/\partial v_j$, $1\le j\le n$.
Note that this is a formal Fourier transform combined with the \emph{Wick
ordering}. 
The Berezin transform extends this by taking the inner
product with a coherent state $\psi_w$. 
\end{observation}

The following notion is very useful.

\begin{definition}\rm
The \emph{Leibniz function} is a map ${\cal C}\times {\cal C} \to {\bf C}$ 
defined as the inner product of the coherent states:
$$ 
           \Upsilon_{wv} = \langle \psi_w,\psi_v \rangle 
$$
for any $v,w$ parametrizing $\cal C$.
\end{definition}

\begin{definition}\rm
The {\em Berezin transform} (the {\em coherent state representation}) 
is defined for an operator $Q$ by,
$$
     \langle Q\rangle_{wv} = {\langle \psi_w, Q\psi_v \rangle 
                            \over \langle \psi_w,\psi_v \rangle}\,.
$$ 
\end{definition}

The algebraic raising operators can be expressed by the Leibniz function,
$\displaystyle \langle \hat R_j \rangle_{wv}
=\Upsilon^{-1}\frac{\partial}{\partial v_j}\Upsilon
=\partial(\log\Upsilon)/\partial v_j$. 
Since $L_j$ is adjoint to $R_j$, we can get $L_j$ by differentiating with
respect to $w_j$. Suppose $\Upsilon$ satisfies a system
of first-order partial differential equations 
$$ 
\frac{\partial\Upsilon}{\partial w_j} =
       \check f_j(v,\frac{\partial}{\partial v})\,\Upsilon
$$
for some operator functions $\check f_j$. Then, from the above discussion, 
we see that $\hat L_j$ is given by
$$
       \hat L_j = f_j({\cal R},{\cal V})
$$
Note that the converse holds as well. Namely, if we have $\hat L_j$
expressed via ${\cal R}$ and ${\cal V}$,
then $\Upsilon$ satisfies the corresponding partial differential
equation. In some cases, this can be used to find $\Upsilon$.\\

The final step is to find in our representation $n$ commuting, 
self-adjoint operators $X_j$.  They will generate a unitary group, 
$\exp(i\sum_js_jX_j)$, with $s=(s_1,\ldots,s_n)\in {\bf R}^n$ and
$i=\sqrt{-1}$.
The scalar function defined by
$$
      \phi(s) = \langle \Omega, \exp(i\sum_js_jX_j)\Omega \rangle 
$$
will be required to be positive-definite. 
Then {\it Bochner's Theorem} assures that $\phi(s)$ is the Fourier
transform of a positive measure, which gives the joint spectral
density of the observables $(X_1,\ldots,X_n)$.\\

For so($n,2$), we will identify these as a (multivariate) random variable 
on the Lorentz cone $\{x_1>0,x_1^2>x_2^2+\cdots+x_n^2\}$
in Minkowski space ${\bf R}^{n,1}$.\\

There are several ways to proceed with the outlined plan. 
One way is to study a matrix realization of the Lie algebra. 
Another is to start from the Leibniz function, which has been 
calculated in \cite{BZ75a} and in \cite{Hua}, and reconstruct 
the Lie algebra from it. We will show how both of these approaches work. \\

\section{Matrix version of so($n$,2)}
 
Consider the ``Lorentz'' group SO($n$,2) of transformations of the 
($n+2$)-dimensional real ``Minkowski'' space ${\bf R}^{n,2}$ of 
signature ($n$,2). The two-dimensional ``time'' leads to two sets 
of independent boosts. Besides the spatial rotations, the 
group contains a 1-dimensional subgroup, so(2), of temporal rotations. 
We shall start with the $(n+2)\times(n+2)$ skew-symmetric matrices 
$\rho_{kl}=E_{kl}-E_{lk}$, for $1\le k,l\le 
n+2$, where $E_{ij}$ denotes the matrix 
consisting of zeros except for 1 at the $(ij)$-entry, .

\begin{notation}\rm
In the following, indices $j$ and $k$ run from $1$ to $n$, referring to $n$ 
``spatial coordinates.'' Subscripts $n+1$, $n+2$ refer to ``time
coordinates.''\\
\end{notation}

%
%

One defines and checks that

\begin{proposition}\label{prop:basismatr}
  The operators $R_j$, $L_j$, for $1\le j\le n$, and $\rho_0$ defined by
  \begin{eqnarray*}
    R_j = \rho_{j,n+2}+i \,\rho_{j,n+1},\quad
    L_j = \rho_{j,n+2}-i \,\rho_{j,n+1},\quad
    \rho_0=2i\,\rho_{n+1,n+2}  
  \end{eqnarray*}
along with $\{\rho_{jk}\}$, $1\le j,k\le n$, 
form a basis of \emph{so($n$,2)} corresponding to a Cartan decomposition as in
equations (\ref{eq:CC}) and (\ref{eq:CC2}).
\end{proposition}

The following relations hold; the root space relations
$$
        [\rho_0,R_j]=2R_j,\quad 
        [L_j,\rho_0]=2L_j,\quad 
        [\rho_{jk},\rho_0]=0
$$
and
\begin{equation}\label{eq:commrels}
        [L_j,R_j]=\rho_0, \quad 
        [L_k,R_j]=2\rho_{jk},\quad 
        [\rho_{jk},L_k]=L_j
\end{equation}
The involution (adjoint map) given by $R_j^*=L_j$ is effectively a complex 
conjugation. The commutation relations determine the involution for the
remaining
elements of $\g$ (since $\mathcal L$ and $\mathcal R$ generate $\g$ as 
a Lie algebra). 
Hence, $\rho_0$, is automatically symmetric, $\rho_0^*=\rho_0$, since it
equals 
a commutator of mutually adjoint elements. 
Also, the $\rho_{jk}$ are skew-symmetric with respect to this involution, 
as follows from relations $2\rho_{jk}=[L_k,R_j]$. \\

Now, we want to find commuting symmetric operators that will provide $n$
commuting 
self-adjoint operators spanning $\g$.  
(Note that even though we have complex numbers in the matrices, we in
fact are using a ``real form'' of $\g$, admitting only real coefficients.)
It turns out that conjugating by $\exp(L_1)$ almost ``does the job." 
More precisely,

\begin{proposition}
The elements
\begin{eqnarray*}
       X_1 &=& 2(\rho_{1,n+2}+i\,\rho_{n+1,n+2})=R_1+L_1+\rho_0 \\
       X_j &=&  2i\,(\rho_{1,j}+i\,\rho_{n+1,j})=-i\,(R_j-L_j-2\rho_{1j}) 
\end{eqnarray*}
for $2\le j\le n$, form a commuting family of 
Hermitian-symmetric elements in $\g$.
\end{proposition}

\begin{proof}
Since the $\mathcal R$ is abelian, conjugating it by a fixed element of the 
group will yield an abelian algebra.
Calculating the adjoint group action $\exp(L_1)R_j\exp(-L_1)$ 
(with a use of commutation relations) yields the indicated operators. 
For $j>1$, the result is skew-symmetric, thus requiring the factor of $-i$ 
for those $X_j$.
\end{proof}

Here are some explicit matrices for $n=3$.
\begin{eqnarray*}
R_1=\pmatrix{ 0&0&0&i&1\cr
              0&0&0&0&0\cr
              0&0&0&0&0\cr
             -i&0&0&0&0\cr
             -1&0&0&0&0\cr} ,
&\quad&
L_1=\pmatrix{ 0&0&0&-i&1\cr
              0&0&0&0&0 \cr
              0&0&0&0&0 \cr
              i&0&0&0&0\cr
             -1&0&0&0&0\cr} \\
\mathstrut\\
X_1=\pmatrix{ 0&0&0&0&2\cr
              0&0&0&0&0\cr
              0&0&0&0&0\cr
              0&0&0&0&2i\cr
             -2&0&0&-2i&0\cr}, 
&\quad&
X_2=\pmatrix{0 &2i&0&0&0\cr
             -2i&0&0&2&0\cr
              0&0&0&0&0\cr
              0&-2&0&0&0\cr
              0&0&0&0&0\cr}
\end{eqnarray*}

and

$$
X_3=\pmatrix{ 0&0&2i&0&0\cr
              0&0&0&0&0\cr
              -2i&0&0&2&0\cr             
              0&0&-2&0&0\cr
              0&0&0&0&0\cr}
\phantom{\pmatrix{ 0&0&-2i&0&0\cr
              0&0&0&0&0\cr
              2i&0&0&2&0\cr             
              0&0&-2&0&0\cr
              0&0&0&0&0\cr}}$$

\begin{remark}\rm
Note that we {\it are\/} using a Hermitian structure here for the inner
product so that multiplication by $i$ converts a skew operator to a
symmetric one.
\end{remark}

\subsection{Representation space}

In the matrix formulation given above, we have a vacuum vector
$$\Omega=\pmatrix{ {\bar 0} \cr 1\cr i}$$
where ${\bar 0}$ stands for a column of $n$ 0's. This vector satisfies
$$
      L_j\Omega=0,        \quad 
      \rho_{jk}\Omega=0,  \quad 
      \rho_0\Omega= -2\Omega
$$
Coherent states can be found readily,
$$ 
     \exp(v_jR_j)\Omega = 
             \pmatrix{ 2i\,{\bf v} \cr 
                             1+v^2 \cr 
                        i\,(1-v^2) \cr}
$$
where ${\bf v}$ is a column vector with components $v_j$ and
$v^2=v_jv_j$.\\

Recalling equation (\ref{eq:grpaction}), this leads to

\begin{proposition}\label{prop:LF}
Let $g$ denote a group element, as in equation (\ref{eq:grpel}),
then we can recover $v$ and $u$ from 
$g\Omega=\pmatrix{{\bf v_0}\cr 
                       a \cr 
                       b \cr }$ by
$$ 
     {\bf v} = \frac{1}{b+ia}\,{\bf v_0},
      \quad
     e^{-2u} = -\half\,\frac{{\bf v_0}^\top{\bf v_0}}{a+ib}
$$
where $\top$ denotes transpose.
\end{proposition}

\subsection{Leibniz function}

Here we calculate the Leibniz function from the matrix representation.
Since $R$'s are adjoint to $L$'s, we have
$$
\Upsilon_{wv} = \langle\, e^{\bf w\cdot R}\Omega, \; 
                              e^{\bf v\cdot R}\Omega\,\rangle 
                  = \langle\,\Omega,\;  
                              e^{\bf w\cdot L}e^{\bf v \cdot
R}\Omega\,\rangle  
$$
where, e.g., ${\bf w\cdot L} = w_jL_j$, and similarly for 
${\bf v \cdot R}$.
As a result we get

\begin{theorem}\label{thm:LF}
  In the matrix realization of so($n$,2) given in Proposition
  \ref{prop:basismatr}, the Leibniz function is
$$
      \Upsilon_{wv}=1-2{\bf w }^\top{\bf v}+w^2v^2
$$
\end{theorem}

\begin{proof}
First compute
\begin{equation}\label{eq:LLF}
      \exp(w_jL_j)\exp(v_jR_j)\Omega
        = \pmatrix{2i({\bf v}-v^2{\bf w })\cr
                    -2{\bf w }^\top{\bf v}+1+v^2+w^2v^2\cr
                   -2i{\bf w }^\top{\bf v}+i\,(1-v^2+w^2v^2)\cr}
\end{equation}

When this is expressed in factored form 
(cf. equations (\ref{eq:grpel}) and (\ref{eq:grpaction})),
taking inner products with $\Omega$ eliminates all factors except for 
$\langle \Omega,e^{\rho_0u}\Omega\rangle$. In general, this is
$e^{\tau\,u}$ with $u$ a function of $v$'s and $w$'s. In the matrix
realization above, applying $\rho_0$ to $\Omega$ shows that $\tau=-2$.
The rest follows from equation (\ref{eq:LLF}) using the result for
$\exp(-2u)$ in Proposition \ref{prop:LF}.

\end{proof}

Generally, we want $\rho_0$ to act on $\Omega$ as multiplication by
$\tau$.
This suggests that for $e^{\tau u}=(e^{-2u})^{-\tau/2}$ we have in general
$\Upsilon_{wv}=(1-2{\bf w }^\top{\bf v}+w^2v^2)^{-\tau/2}$. We can now
check agreement with the  results in \cite{BZ75a,PLV}, cf. the Bergman kernel
function given in \cite{Hua}. 
A main feature of the Leibniz function is that expanded in powers of 
$v$'s and $w$'s it yields the generating function for the inner
products of elements of the basis for the Hilbert space. In general,
there are conditions on the values of $\tau$, the Gindikin set, for which the
Hilbert space has a positive-definite inner product. In this regard,
in addition to Berezin's paper, see \cite{Faraut-Koranyi}.

\subsection{Distribution of the observables}

In the matrix representation, the raising operators are nilpotent, $R_j^3=0$,
and hence $X_j^3=0$ for all $1\le j\le n$. 
Consequently, the exponentials reduce to quadratics and the computations are
very fast. With the vacuum vector as above, we find
$$
        \exp(z_jX_j)\Omega
           =\pmatrix{ 2i(z_1-\zeta^2)\cr 
                                 2z_2\cr 
                               \vdots\cr 
                                 2z_n\cr
                      1-2z_1+2\zeta^2\cr
                            i(1-2z_1)\cr}
$$
where $\displaystyle\zeta^2=z_1^2-\sum_{j\ge 2}z_j^2$. Applying
Proposition \ref{prop:LF}, we have

\begin{proposition}\label{zv}
Let $\displaystyle h^2=(1-z_1)^2-\sum_{j\ge2}z_j^2$.
For a group element generated by the $X_j$ acting on the vacuum, 
$\exp(z_jX_j)\Omega\displaystyle\phantom{\biggm|}$, 
the $v$ and $u$ variables are given according to 
$$ 
     {\bf v}=\frac{1}{h^2}\pmatrix{1-z_1-h^2\cr 
                                       -iz_2\cr
                                      \vdots\cr 
                                       -iz_n\cr},
\quad  \exp(-2u)=h^2
$$
\end{proposition}

With 
$\displaystyle
e^{\tau \,u}=h^{-\tau}=\left((1-z_1)^2-z_2^2-\cdots-z_n^2\right)^{-\tau/2}$
  as the Fourier-Laplace transform of the joint spectral
  density of the $X_j$, we can identify it as a measure on
  the Minkowski cone $\{x_1>0,x_1^2>x_2^2+\cdots+x_n^2\}$. 
  in ${\bf R}^{n,1}$. 
  See, e.g.,   \cite{Letac} as well as the references mentioned above 
  for determining positivity. Up to an exponential factor
  in $x_1$ the density is the {\it Wishart distribution on the Lorentz 
    cone.} See \cite[Chapter XVI]{Faraut-Koranyi} and Casalis
  \cite{C}. The important feature is that the
  positivity implies (means) that we have the Fourier-Laplace transform of
  probability measures which are given by a function raised to a
  power in the Fourier domain. Thus, the measures form a convolution
  family and with a continuous parameter $\tau=t/\hbar$, we have the
  fundamental solution to an evolution equation with generator $u(D)$,
  replacing $(z_1,\ldots,z_n)$ in $u$ as a function of $z$ by
  $(D_1,\ldots,D_n)$, $D_j=d/dx_j$. Thus, $u(D)$ is a `natural
  Hamiltonian' --- generator of time-translations --- associated to the Lie
algebra.\\

Now, we have two expressions, hence coordinate systems, for a
coherent state,

\begin{equation}\label{eq:zandv}
               \exp(z_jX_j)\Omega = e^{\tau u}\exp(v_jR_j)\Omega
\end{equation}

where $u=u(z)$ and $v_j=v_j(z)$ are functions of $z=(z_1,\ldots,z_n)$.
In order to express the basis of the Hilbert space in
terms of the $X$'s rather than $R$'s, we must solve (\ref{eq:zandv})
for the $z_j$ in terms of the $v$'s. 
This will give the \emph{generating function} for the basis
expressed in terms of the $X_j$, written in spectral form as variables
$x_j$.\\

In general,
$$
        e^{v_jR_j}\Omega=\exp\left[x_jz_j(v)-\tau\, u(z(v))\right]
=\sum_{k_1,\ldots,k_n}\frac{v_1^{k_1}\cdots
  v_n^{k_n}}{k_1!\cdots k_n!}\,
\state k_1,\ldots,k_n.
$$
where $z_j=z_j(v)$ are the components of the (functional) inverse to $v(z)$.

\begin{theorem}
The generating function for the basis $\state k_1,\ldots,k_n.$ is
$$
      \exp\left(\frac{x_1(v_1+v^2)+i\,({\bf x}\cdot{\bf
v}-x_1v_1)}{1+2v_1+v^2}\right)\,
             (1+2v_1+v^2)^{-\tau/2}
$$
where $v^2=v_jv_j$ and ${\bf x}\cdot{\bf v}=x_jv_j$.
\end{theorem}

\begin{proof}
To start, note that $e^{-2u}=h^2$ entails $e^{-\tau\,u}=h^{\tau/2}$. Now we
must
solve for the $v$'s in Proposition \ref{zv}. First,
$$ 
        v_1=(h^2-1-z_1)/h^2 \hbox{ implies }
        1+z_1=h^2(1-v_1)
$$
And with $z_k=ih^2v_k$, we square and re-sum on the left-hand side to
yield 
$$
       h^2=h^4\left((1-v_1)^2+\sum_{k\ge2}v_k^2\right)
$$
from which $h^2=(1+2v_1+v^2)^{-1}$. Expressing $h^2$ in terms of $v$'s 
in the above expressions for the $z_j$ yields the result.
\end{proof}

For $n=1$, the terms for $k\ge2$ drop out, reducing to a 
generating function for Laguerre polynomials. Hence, in general, the
basis in the $x$-variables offers a generalization of the classical
Laguerre polynomials.

\section{Leibniz function and the Lie algebra}

Now we shall show how the Lie algebra structure expressed in terms
of the combinatorial raising and lowering operators ${\cal R}_j$ and ${\cal
V}_j$,
the hat-representation, can be constructed from the Leibniz
function. (Recall Remark \ref{rem:hatrep}.)\\

\begin{theorem}
The hat-representation has the form
$$ \hat R_j={\cal R}_j,\quad \hat L_j  
          = \tau {\cal V}_j + 2\left({\cal R}_l{\cal V}_l\right){\cal V}_j 
             -{\cal R}_j{\cal V}^2$$
for the algebraic raising and lowering operators, while the rotation operators
are given by
$$\hat\rho_0=\tau + 2{\cal R}_l{\cal V}_l,\quad
\hat\rho_{jk}={\cal R}_j{\cal V}_k-{\cal R}_k{\cal V}_j$$
\end{theorem}
\begin{proof}
Theorem \ref{thm:LF} provides us the Leibniz function for our
representation of so($n$,2)  
$$
      \Upsilon=(1-2{\bf w}^\top{\bf v}+w^2v^2)^{-\tau/2}
$$
Differentiating, we obtain 
$$
   \frac{1}{\Upsilon}\,\frac{\partial \Upsilon}{\partial w_j}
=
   \tau\,\frac{v_j-w_jv^2}{1-2{\bf w}^\top{\bf
       v}+w^2v^2}\hbox{\quad and\quad }
   \frac{1}{\Upsilon}\,\frac{\partial \Upsilon}{\partial v_j} 
   =
   \tau\,\frac{w_j-v_jw^2}{1-2{\bf w}^\top{\bf v}+w^2v^2} 
$$

Combining these, we find the system of partial differential equations
$$
   \frac{\partial \Upsilon}{\partial w_j}
      =  \tau\, v_j\Upsilon +2v_j \frac{\partial \Upsilon}{\partial v_l}v_l
       - \frac{\partial \Upsilon}{\partial v_j}\,v^2
$$
(implied summation over $l$ in the middle term)
from which we can read off the result stated for $\hat L_j$.
Since $\hat R_j={\cal R}_j$, 
taking the commutator with $\hat L_j$, the first relation in equation
(\ref{eq:commrels}) yields
$
\hat\rho_0 = \tau + 2{\cal R}_l{\cal V}_l
$.
And from the middle relation in equation
(\ref{eq:commrels}),
$$
      [\hat L_k,\hat R_j] = 2({\cal R}_j{\cal V}_k-{\cal R}_k{\cal V}_j)
$$
yields $\hat\rho_{jk}$.
\end{proof}

One can easily check the adjoint action of $\hat\rho_0$ as well as 
the remaining commutation relations corresponding to equation
(\ref{eq:commrels}).
Finally, note that $\hat L_j$ can be written in the form
$$
      \hat L_j=\hat\rho_0{\cal V}_j-{\cal R}_j{\cal V}^2
$$
which is a variation on the Bessel operator.

\section{Conclusion}
An important feature of our approach is the identification of an
interesting abelian subalgebra that provides a family of commuting
observables from the viewpoint of quantization. The special element
$\rho_0$ turns out to be dual to a natural Hamiltonian generating a
convolution semigroup of measures yielding the joint spectral density
of the observables of interest. It is important to note that in
Berezin, e.g., \cite{BZ75a}, what is considered as Planck's constant
should be in fact the ratio $t/\hbar$, namely the ratio between a
time variable and a fixed constant. In this paper, the time in
so($n$,2) is represented by (imaginary) so(2) and in the representation
space by 
a positive real variable, the corresponding weight. The physical
interpretation of this aspect remains to be explored.


\end{document}